\documentclass[12pt,epsf]{article}
\usepackage{amsmath}
\usepackage{amssymb}
\begin{document}
\def\a{\alpha}
\def\b{\beta}
\def\c{\varepsilon}
\def\d{\delta}
\def\e{\epsilon}
\def\f{\phi}
\def\g{\gamma}
\def\h{\theta}
\def\k{\kappa}
\def\l{\lambda}
\def\m{\mu}
\def\n{\nu}
\def\p{\psi}
\def\q{\partial}
\def\r{\rho}
\def\s{\sigma}
\def\t{\tau}
\def\u{\upsilon}
\def\v{\varphi}
\def\w{\omega}
\def\x{\xi}
\def\y{\eta}
\def\z{\zeta}
\def\D{{\mit \Delta}}
\def\G{\Gamma}
\def\H{\Theta}
\def\L{\Lambda}
\def\F{\Phi}
\def\P{\Psi}

\def\S{\Sigma}

\def\o{\over}
\def\beq{\begin{eqnarray}}
\def\eeq{\end{eqnarray}}
\newcommand{\gsim}{ \mathop{}_{\textstyle \sim}^{\textstyle >} }
\newcommand{\lsim}{ \mathop{}_{\textstyle \sim}^{\textstyle <} }
\newcommand{\vev}[1]{ \left\langle {#1} \right\rangle }
\newcommand{\bra}[1]{ \langle {#1} | }
\newcommand{\ket}[1]{ | {#1} \rangle }
\newcommand{\EV}{ {\rm eV} }
\newcommand{\KEV}{ {\rm keV} }
\newcommand{\MEV}{ {\rm MeV} }
\newcommand{\GEV}{ {\rm GeV} }
\newcommand{\TEV}{ {\rm TeV} }
\def\slash#1{\ooalign{\hfil/\hfil\crcr$#1$}}
\def\diag{\mathop{\rm diag}\nolimits}
\def\Spin{\mathop{\rm Spin}}
\def\SO{\mathop{\rm SO}}
\def\O{\mathop{\rm O}}
\def\SU{\mathop{\rm SU}}
\def\U{\mathop{\rm U}}
\def\Sp{\mathop{\rm Sp}}
\def\SL{\mathop{\rm SL}}
\def\tr{\mathop{\rm tr}}

\baselineskip 0.7cm

\begin{titlepage}

\begin{flushright}
UT-KOMABA/07-11
\end{flushright}

\vskip 1.35cm
\begin{center}
{\large \bf
SUSY Unparticle and Conformal Sequestering}
\vskip 2.2cm
Yu. Nakayama${}^{1}$

\vskip 0.5cm

${}^1${\it Institute of Physics, University of Tokyo,\\
      Komaba, Meguro-ku, Tokyo 153-8902, Japan}

\vskip 3.5cm

\abstract{We investigate unparticle physics with supersymmetry (SUSY). The SUSY breaking effects due to the gravity mediation induce soft masses for the SUSY unparticles and hence break the conformal invariance. The unparticle physics observable in near future experiments is only consistent if the SUSY breaking effects from the hidden sector to the standard model sector are dominated by the gauge mediation, or if the SUSY breaking effects to the unparticle sector is sufficiently sequestered. We argue that the natural realization of the latter possibility is the conformal sequestering scenario.}
\end{center}
\end{titlepage}

\setcounter{page}{2}

\section{Introduction}
Recently, there has been a much popularity in unparticle theories \cite{Georgi:2007ek} (see \cite{unpart,unpartHiggs1,unpartHiggs2,SUSYunpart} for further works on the subject), where the unparticle sector shows conformal invariance\footnote{In this paper, we require conformal invariance rather than scale invariance because there is no known single example of scale invariant local field theories that are not conformally invariant. This constraint yields some restrictions on properties of the unparticle sector. See also footnote 2. The author would like to thank H.~Georgi for related discussions.} below a certain high energy scale, and it couples to our standard model sector by higher dimensional interactions. The nontrivial conformally invariant unparticle sector leads to a novel type of observable effects in the standard model sector, which may be accessible in near future experiments at TeV scale.

On the other hand, since one of the most appealing new physics at TeV scale is the supersymmetry (SUSY), it is very natural to consider the supersymmetric extension of the unparticle sector. The first aim of this paper is to investigate the supersymmetric extension of the unparticle sector based on the superconformal field theory. 

Technically, the superconformal field theory in four-dimension \cite{SCFT} is powerful enough to obtain some crucial dynamical information about the unparticle physics. For example, the relation between the $R$-charge and the conformal dimension determines the conformal dimensions of the chiral operators beyond the perturbation theory. We also have severer inequalities for conformal dimensions that are not available in non-supersymmetric theories. In this sense, the introduction of the SUSY in the unparticle sector is theoretically appealing.

Physically, however, the SUSY, in particular the SUSY breaking effect, introduces an extra constraint on the unparticle physics. The original studies on the unparticle physics \cite{Georgi:2007ek} assume that the unparticle sector remains conformal at least down to the weak scale, at which experimental evidence for the unparticle physics is expected. On the other hand, as we will discuss in the main part of this paper, the gravity mediation of the SUSY breaking necessarily gives rise to soft masses for the unparticle sector at the gravitino mass scale. If the gravitino mass is as large as the weak scale, the supersymmetric unparticle physics cannot be realized at the weak scale.

We propose two ways to solve this tension between the SUSY breaking and the unbroken conformal invariance in the unparticle sector. One possibility is to use the gauge mediation \cite{Rattazzi}, where the gravitino mass scale can be low enough to avoid the problem. The other possibility is to tune the Kahler potential so that the gravity mediation does not occur. The latter ``tuning" can be naturally realized through the conformal dynamics by using the conformal sequestering mechanism  \cite{Luty:2001jh,Luty:2001zv} (see also \cite{Dine:2004dv,Ibe:2005pj,Ibe:2005qv,Schmaltz:2006qs,Kachru:2007xp}) between the SUSY breaking hidden sector and the unparticle sector.

The organization of the paper is as follows. In section 2, we present a concrete example of the SUSY unparticle sector based on the SQCD in the conformal window. In section 3, we discuss the conformal symmetry breaking in the unparticle sector. In section 4, we investigate the conformal symmetry breaking induced by the SUSY breaking mediation from the hidden sector. We propose two ways to avoid the high energy breaking of the conformal symmetry by using the gauge mediation or the conformal sequestering. In section 5, we summarize the paper and give some concluding remarks.

\section{SQCD as SUSY unparticle sector}
Our set up is based on three sectors: the first sector is the SUSY standard model sector, the second sector is the SUSY breaking hidden sector, and the last sector is the SUSY unparticle sector. They are weakly interacting with each other through higher dimensional operators.
In this section, we discuss some important properties of the SUSY unparticle sector by presenting a concrete example.

One of the simplest examples of the SUSY unparticle sector is given by the $SU(N_c)$ SQCD with $N_f$ flavors \cite{CGT,unpartHiggs1}, which is a natural SUSY extension of the non-SUSY Banks-Zaks model \cite{Banks:1981nn}. We take $\frac{3}{2}N_c \le N_f \le 3N_c$ so that the unparticle SQCD is in the conformal window. We denote the chiral superfields for the $N_f$ flavors (and their scalar top components) by $Q_i$ and $\bar{Q}_{j}$ ($i,j = 1\dots N_f$). $Q_i$ transforms as a fundamental representation of $SU(N_c)$ and $\bar{Q}_j$ transforms as an anti-fundamental representation.
 
In the unparticle SQCD in the conformal window, we have three possible scalar operators with the ultraviolet (UV) dimension two.\footnote{We do not consider vector or higher tensor operators in this paper. For any {\it conformal} field theories, the conformal dimensions of such operators are bounded from the unitarity as $d\ge 1+j_R+j_L$, where $j_R$ and $j_L$ are $SL(2;\mathbf{C})$ spins \cite{Mack:1975je}, so they are less dominant in the unparticle physics. This bound seems overlooked in some unparticle literatures. For a hypothetical scale invariant field theory that is not conformally invariant, this bound might be violated, and observation of such a violation would be of great interest theoretically. 
} In the unparticle physics, these operators will play dominant roles because they are the lowest dimensional gauge invariant scalar operators in the unparticle sector.

The first one is given by the mesonic (anti-)chiral operators
\begin{eqnarray}
M_{ij} = Q_i\bar{Q}_j \ ,\cr 
M_{ij}^\dagger = Q_i^\dagger\bar{Q}_j^\dagger \ .
\end{eqnarray}
Their conformal dimensions are fixed by the $R$-symmetry:
\begin{eqnarray}
d_{M_{ij}} = 3\frac{N_c-N_f}{N_f} \ .
\end{eqnarray}
We note $1<d_{M_{ij}} < 2 = d_{M_{ij}}^{UV}$.

The second one appears as a scalar component of the conserved current supermultiplets
\begin{eqnarray}
J_{ij} = Q_i^\dagger Q_j \ , \cr
\bar{J}_{ij} = \bar{Q}_i^\dagger \bar{Q}_j \ ,
\end{eqnarray}
where  we only take the $SU(N_f)$ part. Since they are part of the conserved current supermultiplet, they are not renormalized. As a consequence, the dimension of these operators are not renormalized: $d_{J_{ij}} = d_{J_{ij}}^{UV} = 2$.

Similarly the scalar part of the baryon current supermultiplet 
\begin{eqnarray}
J_B = \sum_i Q_i^\dagger Q_i - \bar{Q}_i^\dagger \bar{Q}_i
\end{eqnarray}
is not renormalized as well: $d_{J_{B}} = d_{J_{B}}^{UV} = 2$.

The final one is given by a scalar component of the  Konishi supermultiplet, which is not conserved due to the anomaly:
\begin{eqnarray}
J_A = \sum_i Q_i^\dagger Q_i + \bar{Q}_i^\dagger \bar{Q}_i \ .
\end{eqnarray}
It has a dimension $d_{J_A} = 2 + \gamma_{A}$, where $\gamma_{A} > 0$ is the anomalous dimension of the Konishi supermultiplet.\footnote{The equality $\gamma_{A}>0$ is guaranteed from the superconformal algebra.} This anomalous dimension is related to the slope of the beta function at the fixed point as $\gamma_A = \beta'(\alpha_*)$ (see \cite{Anselmi:1996mq}; the same formula is re-derived in the context of the conformal sequestering in \cite{Luty:2001zv,Ibe:2005pj}).
In the lowest Banks-Zaks approximation, we have $\gamma_{A} = \frac{\left(3-\frac{N_f}{N_c}\right)^2}{3}$.

Even in the strongly coupled regime, we know that the dimensions of $J_{ij}$, $J_{B}$ and $J_A$ are bounded from below by two as a consequence of the superconformal symmetry, so the most relevant operators  to us  in the unparticle physics is given by the (anti-)chiral mesonic operators $M_{ij}$. It is interesting to note that the unparticle SQCD in the conformal window has a Seiberg-dual description, where the UV dimension of $M_{ij}$ is given by $1$ instead of $2$. Since the unparticle physics {\it does} depend on the UV dimension of the operator, we have a chance to distinguish the electric and magnetic UV completions of the low energy effective field theory.

For later purposes, we discuss some properties of the SQCD with soft SUSY breaking terms. As we will see, the SUSY breaking in the hidden sector induces soft parameters in the unparticle SQCD sector as well. We typically have two options.

The first option is that the gaugino in the unparticle SQCD and the squarks in the unparticle SQCD have comparable soft masses.\footnote{We assume that the soft mass squared for the unparticle squarks is positive.} In this case, the low energy effective field theory below the mass scale of the unparticle sector is given by the non-SUSY $SU(N_c)$ QCD with $N_f$ fundamental quarks. The unparticle QCD is supposedly in the confining phase,\footnote{The conformal fixed point would be obtained when $4N_c \lesssim  N_f \le \frac{11}{2}N_c$.  Recent estimations of the lower bound can be found e.g. in \cite{BZfixed}.} and we do not have conformal invariance in the low energy physics. Consequently, we conclude that the unparticle sector shows usual particle physics below the soft mass scale.

The second option is that the unparticle gaugino is much lighter than the unparticle squarks. Then, below the mass scale of the unparticle squarks, we have $SU(N_c)$ QCD with one adjoint Majorana fermion with $N_f$ fundamental quarks. In this case, there is a narrow chance of having a conformally invariant theory (above the unparticle gaugino mass scale) when we take $N_f$ very close to $3N_c$. However, since the lower bound for the conformally invariant QCD is a delicate issue, we will not pursue this possibility further in this paper.

In summary, with the SUSY breaking, the conformal invariance of the unparticle sector is broken at the soft mass scale of the unparticle squarks and gauginos. Below the mass scale, special properties of the unparticle physics is lost and will not be observed experimentally as we will discuss in the following.

\section{Unparticle physics and conformal breaking scale}
We assume that the unparticle sector and our standard model sector interact by the effective operator
\begin{eqnarray}
c\frac{1}{M_\mathcal{U}^{l+d_{UV}-4}} \mathcal{O}_{SM} \mathcal{O}_{UV} \ \label{UVint}
\end{eqnarray}
at the scale $M_{\mathcal{U}}$, where $M_{\mathcal{U}}$ denotes the scale of the ``messenger" between the unparticle sector and the standard model sector. The standard model operator $\mathcal{O}_{SM}$ has a dimension $l$.

Below the scale $\Lambda_U$, where the unparticle sector becomes (approximately) conformally invariant, the interaction \eqref{UVint} can be replaced by
\begin{eqnarray}
c'\frac{\Lambda^{d_{UV}-d}_\mathcal{U}}{M_\mathcal{U}^{l+d_{UV}-4}} \mathcal{O}_{SM} \mathcal{O}_{IR} \ . 
\end{eqnarray}

In general, the unparticle sector cannot remain conformal below the scale of the weak interaction without any further assumption or fine-tuning. This is because we can take $\mathcal{O}_{SM}$ as the Higgs mass operator $|H|^2$ \cite{unpartHiggs1,unpartHiggs2}. Then the conformal invariance of the unparticle sector is spontaneously broken by the vacuum expectation value of the Higgs field: $v^2 = \langle |H|^2\rangle$. The conformal breaking scale is given by \cite{unpartHiggs1}
\begin{eqnarray}
\Lambda_{{\scriptsize {\slash {\cal U}}}} = \left[c'\left(\frac{v}{\Lambda_\mathcal{U}}\right)^2\left(\frac{\Lambda_\mathcal{U}}{M_\mathcal{U}}\right)^{d_{UV}-2}\right]^{\frac{1}{4-d}} \Lambda_\mathcal{U} 
\end{eqnarray}

Unparticle physics can be observed in the energy range $\Lambda_{{\scriptsize {\slash {\cal U}}}} <E<\Lambda_\mathcal{U}$. In order to obtain a reasonable hierarchy between $\Lambda_{{\scriptsize {\slash {\cal U}}}}$ and $\Lambda_\mathcal{U}$, we have to assume either $c'$ or $\frac{v}{\Lambda_\mathcal{U}}$ is small (for $d_{UV}=2$). The latter possibility suggests that the unparticle scale $M_\mathcal{U}$, which is larger than $\Lambda_\mathcal{U}$ from theoretical consistency, is too high to yield any experimentally accessible unparticle effects. Therefore, we have to assume $c'$ is sufficiently small in order to have experimentally observable unparticle physics. Such a possibility can be (technically-)naturally realized by using the (unbroken) SUSY (see also \cite{SUSYunpart}).  Still, we have to face the breaking of the conformal symmetry due to the SUSY breaking and its mediation to the unparticle sector, which will be the main scope of the next section.

\section{Conformal breaking from SUSY breaking}
From the discussion in section 2, if one realizes the SUSY unparticle sector by the SQCD in the conformal window, the soft mass scale $m_0$ for the unparticle SQCD sets another conformally breaking scale $\Lambda_{{\scriptsize {\slash {\cal U}}}} \sim m_0$ in addition to the Higgs mass scale. In order for the unparticle physics to be accessible in near future experiments, the condition $\Lambda_{{\scriptsize {\slash {\cal U}}}} \ll 1\mathrm{TeV} $ should be satisfied. In this section, we investigate the effects of SUSY breaking mediated from the SUSY breaking hidden sector.

First of all, we have to consider (almost) model-independent  universal effects from the anomaly mediation \cite{anomaly}. There is none, however. Since the unparticle sector is conformally invariant, there is no conformal anomaly, and hence there is no soft parameters induced by the anomaly mediation. In the following, we discuss model-dependent effects to the soft masses for the unparticle sector one by one.

Let us begin with the case where the hidden sector SUSY breaking effects are mediated to the standard model sector by the gravity mediation. In this scenario, the gravitino mass $m_{3/2}$ and the standard model sfermion masses $m_0$ are of the same order\footnote{The gaugino mass depends on how $R$-symmetry is broken, but we typically assume that it is also comparable with other SUSY particles.} and given by $m_{0} \sim m_{3/2} \sim \frac{F}{M_{pl}}$, where $F$ denotes the SUSY breaking $F$-term of the hidden sector and $M_{pl}$ denotes the Planck scale.

When the SUSY breaking mediation to the unparticle sector is also given by the gravity mediation, the soft masses for the unparticle SQCD are also given by $\frac{F}{M_{pl}}$. Given that the standard model sfermion masses are above the weak interaction scale, this scenario leads to no near-future-observable unparticle physics. Alternatively, when the SUSY breaking mediation to the unparticle sector is given by the gauge mediation, where the ``messenger" is charged under the unparticle SQCD, the soft mass for the unparticle SQCD is much higher than the gravitino mass. Therefore, this scenario also does not lead to observable unparticle physics.

Let us move on to the case where the hidden sector SUSY breaking effects are mediated to the standard model sector by the gauge mediation. In this scenario, the gravitino mass is much smaller than the standard model sfermion/gaugino masses. 
Furthermore, when the SUSY breaking mediation to the unparticle sector is given by the gravity mediation, the unparticle SQCD acquires soft masses of order of the gravitino mass. Then the light unparticle soft mass is available.  Alternatively, when the SUSY breaking mediation to the unparticle sector is also given by the gauge mediation, the unparticle sector may have a soft mass comparable to the standard model SUSY particles.\footnote{If we assume that the mass scales of the standard model - hidden sector messengers and those of the unparticle - hidden sector are of the same order, the unparticle soft mass tends to be larger than the standard model soft masses due to the large loop factor near the strongly coupled conformal fixed point. However, there is no fundamental reason to take the same masses for messengers living in different sectors, and it is possible to obtain a small soft mass for the unparticle sector.} 

Finally let us consider the case where the hidden sector SUSY breaking effects are mediated to the standard model sector by the anomaly mediation. In this scenario, the gravitino mass is heavier than the standard model sfermion/gaugino masses as $m_{1/2} \sim \frac{\beta_g}{g} m_{3/2}$, where $g$ is the coupling constant of the standard model gauge interaction and $\beta_g$ is its beta function. 

When the SUSY breaking mediation to the unparticle sector is given by the gravity mediation, the unparticle sector obtains soft masses of order of the gravitino mass. Then the unparticle physics near the weak scale is unavailable. The situation with the gauge mediation is even worse, and there would be no observable unparticle physics near future experiments.

In summary, without any fine-tuning of the Kahler potential, only compatible SUSY breaking scenario with the unparticle physics is the gauge mediation between the SUSY breaking hidden sector and the standard model sector. The effect of the SUSY breaking is mediated to the unparticle sector by the gravity mediation.

So far, we have assumed that the Kahler potential is generic (except for the anomaly mediation case, where the direct coupling between the hidden sector and the standard model sector should be sequestered\footnote{We stress, however, that we will focus on the sequestering between the hidden SUSY breaking sector and the unparticle sector. Conceptually the sequestering between the hidden SUSY breaking sector and the standard model sector is an independent issue.}), but since the unparticle sector is in the conformal window, it is natural to study the possibility of the conformal sequestering in the gravity (or anomaly) mediation scenario.

For a success of the sequestering in the coupling between the hidden sector and the unparticle sector, we should demand that the Kahler potential in the Einstein frame is in the sequestered form:
\begin{eqnarray}
K^E=-3\log\left(1-\frac{1}{3}\left(\Phi^\dagger\Phi + S^\dagger S +\sum_i Q_i^\dagger Q_i + \bar{Q}_i^\dagger \bar{Q}_i \right) + \cdots\right) \ , \label{seq}
\end{eqnarray}
where $\Phi$ represents the standard model chiral superfield and $S$ represents the SUSY breaking hidden sector chiral superfield. 

In the conformal sequestering scenario, we first assume that the global symmetry forbids the coupling between the conserved current and the hidden sector superfields such as $S^\dagger S J_{ij}$, $S^\dagger S \bar{J}_{ij}$ and $S^\dagger S {J}_B$ in the (conformal frame) Kahler potential, which can be achieved by demanding $SU(N_f)\times SU(N_f) \times (\text{charge conjugation under } Q_i \leftrightarrow \bar{Q}_i)$  symmetry or their discrete subgroup. This is necessary because the conserved current supermultiplets are not renormalized and the conformal sequestering does not apply for such interactions. We furthermore demand that the interaction such as $S^\dagger S M_{ij}$ must be forbidden e.g. by requiring (discrete) $R$-symmetry. This is because the dimension of $M_{ij}$ is {\it less} than 2, so the conformal dynamics would rather {\it enhance} the coupling.

With these symmetry assumptions, we can see that the remaining interaction $S^\dagger S J_A$ that cannot be forbidden by the symmetry argument is indeed conformally sequestered in our unparticle SQCD model. This is due to the anomalous dimension of the Konishi supermultiplet. More precisely, the coefficient $c$ appearing in the conformal frame Kahler potential\footnote{The Einstein frame Kahler potential and the conformal frame Kahler potential is related by $K^E=-3\log(1-\frac{1}{3}K^c)$.}
\begin{eqnarray}
\delta K^c = c S^\dagger S \left(\sum Q_i^\dagger Q_i + \bar{Q}^\dagger_i \bar{Q}_i\right)
\end{eqnarray}
at the conformal scale $\Lambda_{\mathcal{U}}$ reduces to $c(\Lambda) = c \left(\frac{\Lambda}{\Lambda_{\mathcal{U}}}\right)^{\gamma_A} $ at the scale $\Lambda$.

In this way, for large enough $\gamma_A$, by taking $N_f\sim 2N_c$ in our example, we could obtain a reasonably low energy conformal breaking in the gravity mediation scenario while keeping the unparticle energy scale $M_{\mathcal{U}}$ low enough for the unparticle physics to be observed in near future experiments. In the anomaly mediation scenario, the situation seems still harder because we have to compensate the loop factor as well: we have to take large $\Lambda_{\mathcal{U}}$ and hence $M_{\mathcal{U}}$ for large conformal sequestering and the unparticle physics tends to be unaccessible in future experiments.\footnote{When the conformal dynamics applies also in the SUSY breaking sector, the conformal sequestering could be enhanced and the anomaly mediation scenario would survive.}

\section{Conclusion}
In this paper, we have investigated the SUSY unparticle physics. The SUSY breaking effects induce soft masses in the SUSY unparticle sector, and these masses should be small enough for the unparticle physics to be observed in near future experiments. The gravity mediation, however, typically gives rise to the soft masses of the order of the gravitino mass. 

We have proposed two solutions to this problem. The one is to use the gauge mediation to yield the low scale gravitino mass. The other is to use the conformal sequestering to suppress the effect of the gravity mediation.

In this sense, the observation of the unparticle physics together with  the discovery of SUSY will highly restrict possible mediation scenarios of the SUSY breaking, which could be regarded as a probe of the high energy physics beyond the TeV scale. It would be very interesting to pursue this direction to invent a novel probe for the SUSY breaking pattern. It would be also intriguing to realize the SUSY unparticle sector in the superstring theory. These issues are left for future studies.

\section*{Acknowledgements}
The author would like to thank members of Komaba particle theory group and in particular J.~Yasuda for bringing the author's attention to unparticle physics.
He thanks H.~Georgi for the discussion on scale invariance and conformal invariance.
He also acknowledges the Japan Society for the Promotion of Science for financial support.

\end{document}